\documentclass{emulateapj}
\usepackage{color}

\newcommand{\feoh}{[{\rm Fe} / {\rm H}]}
\newcommand{\msun}{\, M_\odot}
\newcommand{\Zsun}{\, Z_\odot}
\newcommand{\mmd}{M_{\rm md}}
\newcommand{\dm}{\Delta_{\rm M}}
\newcommand{\Tv}{T_{\rm vir}}
\newcommand{\cempnos}{CEMP-no$s$}
\newcommand{\cemps}{CEMP-$s$}
\newcommand{\alphafe}{\abra{\alpha}{Fe}}

\def\abra#1#2{[{\rm #1}/ {\rm #2}]}

\shorttitle{Chemical Evolution with High-Mass IMF}
\shortauthors{Komiya}

\begin{document}
\title{Extremely Metal-Poor Stars and a Hierarchical Chemical Evolution Model}

\author{Yutaka Komiya\altaffilmark{1}}
\altaffiltext{1}{National Astronomical Observatory of Japan, Osawa, Mitaka, Tokyo, Japan}

\begin{abstract}

Early phases of the chemical evolution and formation history of extremely metal poor (EMP) stars are investigated using hierarchical galaxy formation models. 
We build a merger tree of the Galaxy according to the extended Press-Schechter theory. 
We follow the chemical evolution along the tree, and compare the model results to the metallicity distribution function (MDF) and abundance ratio distribution of the Milky Way halo. 
We adopt three different initial mass functions (IMFs). 
In a previous studies, we argue that typical mass of EMP stars should be high-mass($\sim10\msun$) based on studies of binary origin carbon-rich EMP stars. 
In this study, we show that only the high-mass IMF can explain a observed small number of EMP stars. 
For relative element abundances, the high-mass IMF and the Salpeter IMF predict similar distributions. 
We also investigate dependence on nucleosynthetic yields of supernovae (SNe). 
The theoretical SN yields by Kobayashi et al.(2006) and Chieffi \& Limonge (2004) show reasonable agreement with observations for $\alpha$-elements. 
Our model predicts significant scatter of element abundances at $\feoh<-3$. 
Best fit yields for one zone chemical evolution model by Francois et al.(2004) well reproduces the trend of the typical abundances of EMP stars but our model with their yield predicts much larger scatter of abundances than the observations. 
The model with hypernovae predicts Zn abundance in agreement with observations but other models predict lower $\abra{Zn}{Fe}$. 
Ejecta from the hypernovae with large explosion energy is mixed in large mass and decreases scatter of the element abundances. 
\end{abstract}


\section{Introduction}
Extremely metal-poor (EMP, $\feoh<-2.5$ in this paper) stars are stars formed in the early universe in terms of chemical evolution. 
They are thought to have been formed at high redshift but are still shining with the glow of nuclear burning in the Local Group. 
Recent large-scaled surveys have identified hundreds of the EMP stars in the Milky Way halo (HK survey, Beers et al.\ 1992: Hamburg/ESO [HES] survey, Christlieb et al.\ 2003, 2008: Sloan Extension for Galactic Understanding and Exploration [SEGUE], Yanny et al.\ 2009). 
Element abundances of these stars are revealed by follow-up spectroscopic observations. 
They provide a means of probing the earliest phases of the evolution of the Milky Way and supernovae (SNe) in the early universe. 

In this paper, we refer to stars with $\feoh\leq -2.5$ as EMP stars, although ``EMP star'' is usually used for $\feoh < -3$ \citep{Beers05}. 
As shown in our previous study \citep[][hereafter Paper I]{Komiya10}, stars with $\feoh\lesssim-2.5$ show some observational and theoretical peculiarities distinguishing them from more metal rich Population~II stars (see Section 2.1 of Paper I). 
Especially, in previous studies \citep{Komiya07,Komiya09}, we show that the initial mass function (IMF) of stars with $\feoh\leq -2.5$ should differ from metal rich stars. 
Typical mass of the EMP stars is $\sim 10 \msun$ and present EMP stars in the Milky Way halo are the low mass minorities. 
We refer to the mother stellar population with $\feoh\leq -2.5$ as EMP population and low mass survivors with nuclear burning now as EMP survivors. 
Stars with $\feoh<-5$ are referred to as hyper metal-poor (HMP) stars. 
 2 HMP stars detected in the Milky Way halo are the most metal deficient objects observed yet (HE1327-2326, Frebel et al.\ 2005: HE0107-5240, Christlieb et al.\ 2002). 

Formation environment of the EMP population stars are thought to differ from present galaxies. 
EMP stars are formed in the process of galaxy formation in the early universe. 
In the $\Lambda$ cold dark matter (CDM) universe, large galaxies like the Milky Way are formed through merger of smaller galaxies as building blocks. 
According to the hierarchical structure formation scenario, a stellar halo is an aggregation of stars formed in the many small galaxies \citep{Searle78, Helmi08}. 
Earlier theoretical studies show that the first stars are formed in very low mass halos with $\sim 10^6\msun$ \citep{Tegmark97, Nishi99} and host galaxies of the second generation of stars are also small \citep{Ricotti02, Wise08}. 
Chemical abundances of these building blocks can differ from each other. 
Unlike metal rich stars, metals in EMP stars are synthesized by only one or a few precursory SN(e). 
Element abundances of the EMP stars can reflect individual characteristics of the precursory SN(e) and their host galaxies. 
A semi-analytic hierarchical approach can provide a framework within which to study the earliest phases of the chemical evolution and the formation history of the EMP stars. 

One important point at issue for the earliest phases of chemical evolution is a possible difference in the IMF of EMP stars \citep[e.g.][]{Abia01, Komiya07}. 
Theoretically, typical mass of stars is large in extremely metal-poor environment and/or in the early universe. 
Numerical simulations show that Population~III stars without metal are very massive \citep[e.g.][]{Bromm99, Abel02}. 
For EMP stars with a little metal, the IMF can be different from nearby stars, too \citep{Omukai05}. 
Existence of the low mass EMP survivors in the Galactic halo proves that some low mass EMP stars can be formed, but typical mass of the EMP population stars can be more massive than Population~I stars. 
\citet{Komiya07, Komiya09} give constraints on the IMF of EMP stars from statistics of observed EMP survivors. 
It is known that large fraction ($\sim 20\%$) of the EMP survivors comprise carbon enhanced stars referred to as Carbon Enhanced Metal-Poor (CEMP) stars. 
More than half of them show large {\it s}-process element enhancement. 
Abundance anomalies of the {\it s}-process element enhanced CEMP stars (\cemps\ stars) are due to binary mass transfer. 
Intermediate massive EMP stars with $0.8-3 \msun$ synthesize carbon and {\it s}-process elements in the asymptotic giant branch (AGB) phase, and pollute their companions to make them \cemps\ stars. 
We propose that CEMP stars without {\it s}-process enhancement (\cempnos\ stars) are also formed through binary mass transfer but from more massive primaries with $4-6\msun$. 
From observational statistics of the \cemps\ and \cempnos\ stars, we give constraints on mass distribution of primary stars of EMP binaries and conclude that typical mass of EMP stars is large \citep{Komiya07}. 
\citet{Komiya09} discuss the constraints of the IMF from CEMP stars again in detail and give an additional constraint from the total number of EMP survivors. 
The number of EMP survivors in the Galactic halo is very small and it indicates that the fraction of low mass survivors among the EMP population is small. 
As a result, a lognormal IMF with medium mass, $\mmd \sim 10\msun$, and dispersion, $\dm \sim 0.4$, can satisfy all the constraints. 
\citet{Lucatello05} also give a constraint on the IMF of EMP stars from statistics of \cemps\ stars and argue that the typical mass of EMP stars is slightly higher than more metal rich stars. 
However, because they do not take account of \cempnos\ stars and assume a stellar evolution model different from ours, they conclude lower typical mass, $\mmd=0.79\msun$. 

In paper I, we build a merger tree of the Galaxy and compute the enrichment history of iron abundance along the tree. 
The high mass IMF with $\mmd=10\msun$ \citep{Komiya07} is adopted in the computations. 
We also discussed origin of HMP stars considering the effect of surface pollution by accretion of metal enriched interstellar matter. 
In this paper, we investigate chemical evolution of several elements with detailed theoretical nucleosynthetic yields of metal deficient massive stars. 
Model results with high and low mass IMFs are compared with compiled observational data. 
To deal with individual characteristics of the EMP stars, all the individual EMP population stars are registered in our computations. 
We discuss not only averaged abundances but also dispersion of them. 

\citet{Tumlinson06} presents a semianalytic hierarchical model for the Galactic halo. 
\citet{Salvadori06} also calculate a hierarchical model with gas blowout from halos by bursty star formation taken into account, but they do not deal with individual stars and do not investigate diversity of the element abundances. 
These previous studies assume the Salpeter IMF for EMP stars. 
Additionally, these previous studies with hierarchical models investigate only iron abundance. 
We calculate formation and evolution of stellar halo with different IMFs and compare the predicted metallicity distribution functions (MDFs) and abundance ratio distributions. 
We use four different sets of core-collapse SN yields calculated for metal-poor stars. 
\citet{Argast00} and \citet{Karlsson05} investigate element abundances of EMP stars using inhomogeneous chemical evolution models. 
However, they do not take account of the merging history of the Galaxy and stars are assumed to be formed randomly in space. 
They also do not investigate the IMF dependence. 

This paper is organized as follows. 
Computation method and assumptions appear in the next section. 
Especially, assumptions about IMFs and SN yields are described in Sections \ref{IMFS} and \ref{yieldS}, respectively. 
In Section~\ref{obsS}, observational sample for comparison is described. 
We give results in Section~\ref{resultS} and conclude the paper in Section~\ref{concS}.

\section{Computation Method}
\subsection{A Hierarchical Chemical Evolution Model for EMP stars}
A hierarchical chemical evolution model for EMP stars is developed in Paper I.  
We built a merger tree semi-analytically using the method of \citet{SK99} based on the extended Press-Schechter formalism \citep{LC93}. 
Along the merger tree, chemical enrichment and formation history of EMP stars are calculated. 
In this paper, abundances of O, Na, Mg, Si, Cr, Fe, and Zn are computed and compared to observations of metal poor stars.  

Stars are assumed to be formed in halos with virial temperature, $\Tv$, higher than $10^3$K. 
Star formation efficiency assumed to be constant and determined to give $\feoh=0$ at $z=0$ ($2.1\times 10^{-11} - 1.1\times 10^{-10}/{\rm yr}$ for the following computations). 
To investigate diversity of element abundances, all the individual EMP population stars are registered in our computations. 
Mass of each EMP star is specified randomly according to the statistical weight with the IMF. 
Half of the all stellar systems are taken to be binaries and a flat mass ratio distribution is assumed. 
Adopted IMFs are described in Section~\ref{IMFS} in detail. 
We set nucleosynthesis yields of each massive star as a function of it's initial mass and metallicity. 
Assumptions about yields are described in Section~\ref{yieldS}. 
Each minihalo is assumed to be chemically homogeneous.  

We use the same assumptions as Model P in Paper~I about radiative and dynamical feedback from massive stars. 
Massive stars ionize the ambient matter. 
At $z<20$, Lyman-Werner background radiation inhibits star formation in mini-halos those are not pre-ionized. 
Energetic SN explosions blowout gas in their host minihalos when their effective kinetic energy, $\epsilon E_{\rm k}$, is larger than binding energy, $E_{\rm bin}$, of the gas of their host halos. 
Gas and metal ejected from minihalos are mixed immediately and homogeneously throughout the intergalactic medium (IGM). 
Assumed explosion energies, $E_k$, of SNe are described in Sec.~\ref{yieldS}. 
From larger halos, a little fraction, $\eta\epsilon E_k/E_{\rm bin}$, of metal ejected by SNe is assumed to go to IGM, where $\epsilon$ is the fraction of supernovae explosion energy converted into kinetic energy, and $\eta$ is the fraction of the input kinetic energy that is retained by gas that escapes their host halo by wind. 
We assume $\epsilon=0.1$ and $\eta=0.1$, but results in this paper are almost independent from these parameters. 
Because of reionization, mini-halos with $\Tv<10^4 {\rm K}$ cannot accumulate gas at $z<10$. 

Computations using different IMFs and different nucleosynthetic yields are presented. 
Adopted IMFs and SN yields are described in following subsections and summarized in Table~1. 

\subsection{IMF}\label{IMFS}
As stated in Sec.1, in our earlier studies, we give constraints on the IMF of the EMP population stars from statistics of EMP survivors. 
We assume lognormal form, 
\begin{equation}
\xi (m)\propto \frac{1}{m} \exp \left[ -\frac{ \{ \log (m/ \mmd) \} ^2}{2\times \dm^2} \right],
\label{eq:IMF-lognormal}
\end{equation}
 and conclude that the high mass IMF with $\mmd=10\msun, \dm=0.4$ is in agreement with all the observational features. 
We use this high mass IMF as the fiducial one. 

\begin{figure}
\plotone{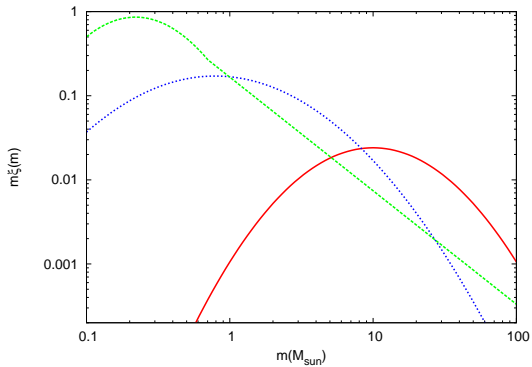}
\caption{Initial mass functions (IMFs) adopted in our computations. 
The solid red, dashed green and dotted blue lines denote IMFs of \citet{Komiya07, Chabrier03} and \citet{Lucatello05}, respectively. 
}
\label{IMFs}
\end{figure}

Based on recent observations, \citet{Chabrier03} presents an IMF of the Galactic spheroid; 
a lognormal IMF with $\mmd=0.22\msun, \dm=0.33$ for stars with $m<0.7\msun$, 
and $\xi(m) \propto m^{-2.35}$ for higher mass stars. 
For comparison, computation with this standard low mass IMF is also presented (Model CK). 
Since this IMF is the same as the Salpeter IMF for high mass and intermediate massive stars, chemical evolution under this IMF is almost the same as the Salpeter IMF.  

The IMF derived by \citet{Lucatello05} is also tested (Model LK). 
They argue a lognormal IMF peaked with slightly higher mass, $\mmd=0.79\msun, \dm=0.51$, based on the statistics of the \cemps\ stars. 

For simplicity, we use the same IMF for all the stellar populations in each computation. 
But we note that the high mass IMF is derived from statistics of EMP survivors. 
In the Galaxy today, more metal rich stars are formed under a low mass IMF. 

The IMF of the first stars can be different from EMP stars. 
In Paper I, we have discussed typical mass of the first stars from viewpoint of the number of surviving Population~III stars.
Observational scarcity of stars with $\feoh<-4$ suggests that the typical mass of the first stars in each mini-halo is higher than the subsequent generations of stars. 
In this paper, $\mmd=40\msun$ is assumed for local first stars without SN progenitor in their host mini-halos. 
The number of the predicted HMP stars becomes comparable with observations under this assumption. 

Many numerical studies for the Population~III star formation argue that one very massive single star (or binary, \citet{Turk09}) with $m>100\msun$ is formed at the center of the primordial mini-halo \citep[e.g.][]{Oshea07, Yoshida08}. 
On the other hand, some studies show that zero-metallicity gas can fragment into multiple pieces and some low-mass stars are also formed without metal\citep{Nakamura01, Clark08}. 
Also in the latter studies, however, the typical mass of Population~III stars is higher than nearby stars. 
\citet{Yoshida07} shows that stars with $\sim 40\msun$ are formed in photo-ionized halo without metal. 
When a Population~III star with $\sim200\msun$ is formed, it explodes as pair instability SN(PISN). 
We discuss the effect of the very massive first stars and PISNe in Appendix. 

\subsection{Supernova Yields}\label{yieldS}
\begin{figure}
\plotone{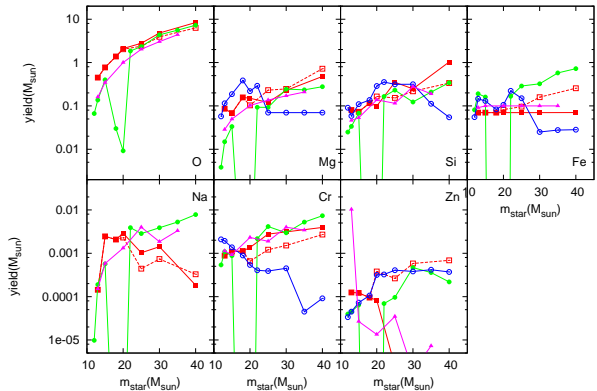}
\caption{
Theoretical SNe~II yields of the zero metallicity massive stars against initial stellar mass by \citet{Kobayashi06}(red squares), \citet{Woosley95}(green filled circles), \citet{Francois04}(blue open circles), and \citet{Chieffi04}(magenta triangles). 
The open squares with dotted lines denote hypernovae yields. 
}
\label{yield}
\end{figure}

In this study, stars with $10-50\msun$ are assumed to explode as Type II SNe (SNe~II). 
In Paper I, we simply assume that any SNe~II eject $0.07\msun$ of iron, but in this study, we consider dependence of yields on stellar initial mass and initial metallicity. 
For SNe~II, we adopt four sets of yields. 
Figure~\ref{yield} summarizes yields of SNe~II with $Z=0$ against initial mass, by theoretical models adopted in this study. 

Theoretical yields by \citet{Kobayashi06} are used as the fiducial one (Model KK). 
They give yields of massive stars with various initial mass and initial metallicity by numerical calculations of stellar evolution and explosive nucleosynthesis at SNe~II. 
They present yields of hypernovae with larger explosion energy, too. 
Following \citet{Kobayashi06}, in this paper, $50\%$ of stars with $m>20\msun$ are assumed to explode as hypernovae. 
Explosion energy of the SNe~II are assumed to be $E_{\rm k}=10^{51} {\rm erg}$ for normal SNe and $E_{\rm k}=10^{51}\times (m/\msun-10) {\rm erg}$ for hypernovae. 
The most prominent nucleosynthetic feature of the hypernova is a large Zn yield. 
We note that, in computation by \citet{Kobayashi06}, iron yields for normal SNe are tuned to $0.07\msun$ by choosing the ``mass cut'' parameter. 
In such computations for SN nucleosynthesis, the location of the boundary between the part of the star that eventually collapses to a compact object and that which is expelled outward is a free parameter and referred to as mass cut. 
An amount of the ejected iron strongly depends on this parameter. 
For hypernovae, parameters involved in the mixing and fallback are determined to give $\abra{O}{Fe} =0.5$. 
We also present a chemical evolution model without hypernova (Model KKn) to show the contribution from hypernovae. 

The theoretical yields by \citet{Woosley95} are also adopted (Model KW). 
We use the explosion models labeled as 12A-22A, 25B, 30C, 35C and 40C in \citet{Woosley95}. 
For explosion energy, resultant kinematic energy of their computations is used ($1.1-3.01\times10^{51}$erg). 

\citet{Francois04} modify the yields by \citet{Woosley95} and suggest the best fit SN yields for $\alpha$-elements and iron group elements, based on the comparison between one zone chemical evolution calculation and observational data. 
For oxygen, they adopt the yield by \citet{Woosley95} without modification. 
Their best fit yields are, as a matter of course, consistent with observations as far as one zone model with the Salpeter IMF goes. 
We adopt their yields to the hierarchical evolution model with the high mass IMF (Model KF). 
Since they do not discuss metallicity dependence of the SN yields, the same yields are assumed for any metallicity in Model KF. 
For Na, we assume the same yield as \citet{Woosley95} since they do not derive best fit one. 

\citet{Chieffi04} also give explosive yields of massive stars from $Z=0$ to $Z=\Zsun$. 
Chemical evolution computation using their results is also presented (Model KC). 
We note that they chose a mass cut parameter to eject $0.1\msun$ of iron for all SNe~II. 
Explosion energy is assumed to $10^51$erg. 

For Type Ia SNe (SNe Ia), a single degenerate scenario is assumed. 
Mass range of a primaries to be SN~Ia is assumed to $2\msun < m < M_{up}$. 
Following the \citet{Greggio05}, $9\%$ of binaries in this mass range become SNe Ia. 
Delay time from star formation to SN explosion is equal to a lifetime of secondary companion of binary. 
Because individual stars are registered in our computation, delay time of each SN Ia is calculated from mass of the secondary star, without assuming a delay time distribution function. 
Event rate and delay time distribution of SN Ia depend on the IMFs. 
We use the W7 model in \citet{Iwamoto99} for metal yields of SNe Ia. 
In this study, we assume frequency and yields of SNe Ia are metallicity independent, for simplicity. 
It is worth noting that some studies argue that the fraction of stars to be SNe Ia depends on the metallicity. 
\citet{Kobayashi98} argue that only stars with metallicity larger than $\feoh>-1$ become SNe Ia. 
But the dependence on the metallicity has not yet been well understood. 

Some stars with mass around $10\msun$ are thought to become electron capture SNe of progenitor AGB stars with an O-Ne-Mg core. 
\citet{Wanajo09} gives nucleosynthesys yields in this type of explosion. 
Amounts of the ejected iron and $\alpha$-elements are much smaller than SNe~II. 
One prominent feature of the predicted yields by \citet{Wanajo09} is a large yield of Zn. 
He argues that O-Ne-Mg SNe can be a main source of Zn in the universe. 
Mass range and fraction of the stars to become O-Ne-Mg SNe are not well revealed. 
We assume that stars with $9-10\msun$ become O-Ne-Mg SNe. 

In this study, we omit the metal ejected by mass loss from intermediate massive stars. 
Because EMP stars are formed in the very early stages of the chemical evolution, metal ejected from the intermediate massive stars which have longer lifetimes than massive stars should be negligible. 
Additionally, the amount of $\alpha$-elements, iron group elements, and Zn provided by the intermediate massive stars are thought to be less than by SNe. 

\begin{table*}
\begin{center}
\caption{Models}
\label{Tmodel}

\begin{tabular}{l|l|l}
\hline

name	& IMF & SN Yield\\
\hline
KK	& \citet{Komiya07}($\mmd=10\msun$) & \citet{Kobayashi06} \\
LK	& \citet{Lucatello05}($\mmd=0.79\msun$) & \citet{Kobayashi06} \\
CK	& \citet{Chabrier03}($\mmd=0.22\msun$) & \citet{Kobayashi06} \\
& & \\
KW	& \citet{Komiya07} & \citet{Woosley95} \\
KF	& \citet{Komiya07} & \citet{Francois04}	\\
KC	& \citet{Komiya07} & \citet{Chieffi04}	\\
KKn	& \citet{Komiya07} & \citet{Kobayashi06}(no hypernova) \\ 
\hline
\end{tabular}
\end{center}
\end{table*}

\section{Observational data}\label{obsS}

We adopt a bias-corrected halo MDF constructed from subsamples of the HES survey with moderate-resolution spectroscopy by \citet{Schorck09}. 
At $\feoh<-3$, they have replaced the moderate resolution values with those derived from high resolution spectroscopy, where available. 
Since the HES survey is biased toward low metallicity, they evaluate biases and give the selected fraction, $f_{s}$, as a function of metallicity and color. 
Instead of the raw MDF, $N_{obs}(\feoh)$,  of the HDS sample, a bias-collected MDF, $N_{obs}(\feoh)/f_{s}(\feoh)$, is plotted to comparison with the model results. 

We also show a MDF compiled by SAGA (Stellar Abundance for Galactic Archaeology) database \citep{Suda08} since we are interested in the low-metallicity tail of the MDF. 
SAGA compiles elemental abundances of EMP halo stars which received high-resolution spectroscopic observations. 
At $\feoh<-3$, metallicity derived by moderate-resolution spectroscopy can be significantly different with that derived by high-resolution spectroscopy. 
High-resolution data tell us accurate metallicity of EMP stars and SAGA contains a large sample of EMP stars. 
We plot raw data compiled by the SAGA database and this is strongly biased toward low metallicity. 
However, we may well regard the selection of target stars for the follow-up observations as being hardly biased below the metallicity of $\feoh \simeq -3$. 
The MDF of \citet{Schorck09} contains only two stars at $\feoh<-3.6$, and none at $\feoh<-4.3$. 
However, currently 19 stars with $\feoh<-3.6$ have been identified in the SAGA sample and two stars have $\feoh<-5$. 

We consider not only the form of the MDF but also the total number of EMP survivors. 
In Paper~I, we have estimated efficiency of the identification of EMP survivors by the HES survey from its coverage area.  
The effective survey area is $S = 6726 \hbox{ deg}^2$ and roughly $40\%$ of the candidates, selected by the objective-prism survey, have been examined by the spectroscopic follow-up observations with medium resolution \citep{Christlieb08}. 
We assume that almost all giant EMP survivors present in the survey fields are detected because of the large limiting magnitude of the HES survey ($B <17.5$). 
When uniform distribution of the stellar halo is assumed, $\sim 5\%$ of the giant EMP survivors in the Milky Way halo are expected to be detected. 
In the following figures, we plot the predicted number of stars which are expected to be in the HES sample, assuming a uniform stellar halo. 
When de Vaucouleurs density distribution is assumed, detection frequency decreases by a factor of 5, since many undetectable stars should be distributed in the inner part of the Galactic halo from the solar orbit.

\begin{figure*}
\plotone{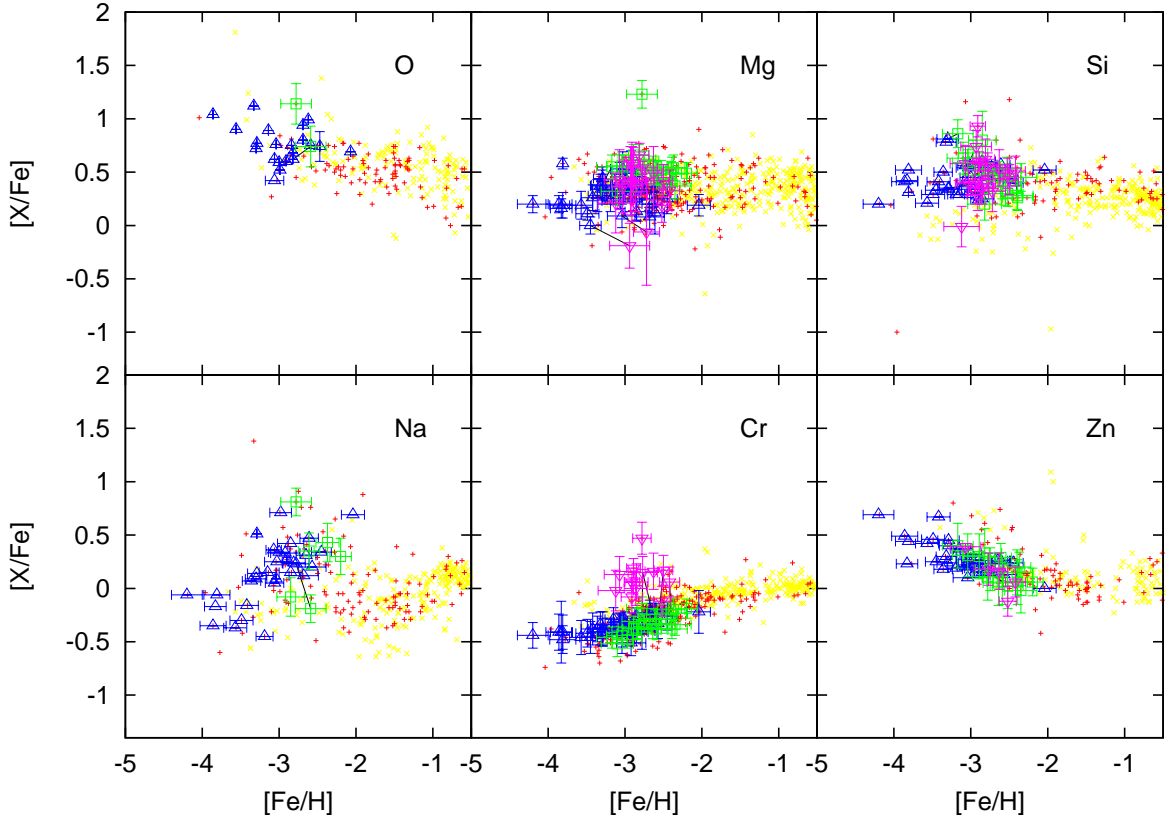}
\caption{Abundance ratio distributions of metal poor stars compiled by the SAGA database. 
Among the SAGA sample, 
stars analysed by the First Stars project ({\it blue triangles}, $\triangle$),  
stars analysed by Aoki et al., ({\it green squares}, $\square$), and
stars in Honda et al.(2004, 2007) ({\it magenta inverted triangles}, $\triangledown$) are plotted with error bars, where available. 
Other giants ({\it red plus signs}, $+$) and main sequence stars ({\it yellow crosses}, $\times$) in the SAGA sample are plotted without error bars. 
Stars in common between two or three subsamples are connected with lines. 
See text in detail. 
}
\label{XFeobs}
\end{figure*}

For elemental abundance ratio, we adopt data compiled by the SAGA database\citep{Suda10}. 
Figure~\ref{XFeobs} shows distributions of O, Na, Mg, Si, Cr, and Zn abundances relative to iron against $\feoh$. 
We select high-resolution sample of $R\geq20000$. 
We note that, since SAGA compiles data from many sources, scatter of the abundances by the SAGA dataset can be larger than the intrinsic scatter of the stellar abundances. 
  To see the systematic difference between literatures, three subsamples analysed by different authors are plotted by different symbols with error bars.
  Blue triangles$(\triangle)$ in Fig.~\ref{XFeobs} denote data analysed by the First Stars project (\citet{Hill02} and other 13 papers of the series).
  Green squares($\square$) denote data of which first author of source paper is W.Aoki (\citet{Aoki02} and other 15 papers in entry list of SAGA). 
  Magenta inverted triangle $(\triangledown)$ shows sample of \citet{Honda04, Honda07}.
  These three subsamples are analysed assuming a plane parallel stellar atmosphere (1D) and local thermal equilibrium (LTE) but
  there are systematic difference in the abundance ratio owing to difference in model atmospheres, parameters used during the analysis, and lines used in analysis. 
  Other stars are plotted with red plus signs ($+$) and yellow crosses ($\times$) for giants($T_{\rm eff}\leq 6000$ and $\log g\leq 3.5$) and dwarfs, respectively. 
  
  We plot only giant stars in the following figures since there is systematic shift in stellar abundances between giants and dwarfs for some elements \citep{Bonifacio09}. 
  Carbon enhanced stars with $\abra{C}{Fe}>0.5$ are not plotted because their surface is thought to be polluted by binary mass transfer. 
\citet{Komiya07} show that carbon on these stars originates in the intermediate massive companion stars and their surface abundances of O, Na and Mg are also affected by binary mass transfer \citep{Nishimura09}. 
  Recently, abundance determinations with a non-LTE scheme\citep[e.g.][]{Mashonkina08,Andrievsky10} or with 3D model atmospheres\citep[e.g.][]{Asplund01,Gonzalez08} are carried out. 
  Difference of abundances determined with these models from 1D LTE models is often of order $\sim 0.5$ dex. 
  They possibly diminish the shift between giant and turnoff stars and decrease dispersion of the abundance distribution. 
  We comment about 3D/nonLTE effects for some elements in next section with comparison to model results.

\section{Results and Discussion}\label{resultS}
\subsection{Metallicity Distribution Functions}
\begin{figure}
\plotone{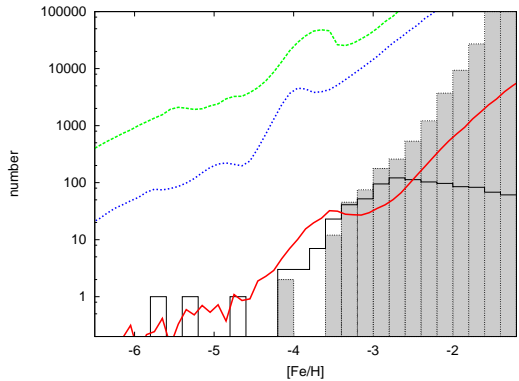}
\caption{
Predicted MDFs assuming three different IMFs. 
The solid red, dashed green, and dotted blue lines show resultant MDFs of Models KK, LK, and CK, respectively. 
The gray histogram denotes the bias corrected observational MDF by the HES survey\citep{Schorck09}. 
Black solid line shows the raw MDF compiled by the SAGA database. 
The predicted MDFs had been normalized in consideration of the survey volume of the HES survey. 
Only the Model KK with the high mass IMF by \citet{Komiya07} is consistent with the observed total number of EMP stars. 
}
\label{IMFMDF}
\end{figure}
 
Figure~\ref{IMFMDF} shows resultant metallicity distribution functions (MDFs) for three models using different IMFs. 
Solid red, dashed green and dotted blue lines denote Models KK, LK, and CK, respectively. 
All three models predict similar patterns of the MDFs but quite different total number of EMP survivors. 
This is because fractions of low-mass stars are different. 
As seen in Fig.~\ref{IMFMDF}, Model KK with the high mass IMF is consistent with observations but other models with the lower mass IMFs predict many more EMP survivors. 
We may overestimate the efficiency of the identification of EMP survivors by the HES survey, because we assume homogeneity of the Galactic stellar halo and the HES survey reaches the outer end of the Galactic halo. 
However, as seen in Fig.~\ref{IMFMDF}, the predicted number of EMP survivors for Models LK and CK is $\sim 100-1000$ times larger than observations and this large discrepancy cannot be explained by spatial inhomogeneity of the stellar halo and/or insufficiency of the survey. 
This indicates that typical mass of EMP population stars is significantly higher than nearby Pop.~I stars, as shown in our earlier studies. 

\begin{figure}
\plotone{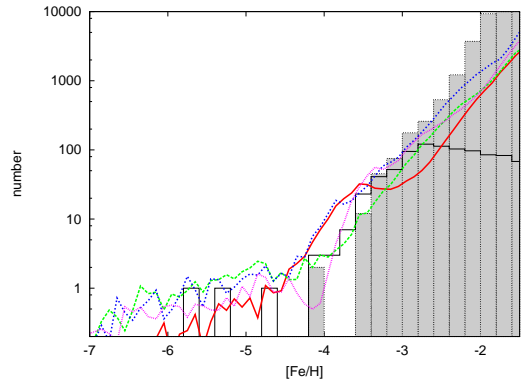}
\caption{
Same as Figure~\ref{IMFMDF} but for models with different SN yields. 
The solid red, long-dashed green, short-dashed blue and dotted magenta lines show results of Models KK, KW, KF, and KC, respectively.
}
\label{yieldMDF}
\end{figure}

Figure~\ref{yieldMDF} shows dependence on the SN yields. 
Solid red, long-dashed green, short-dashed blue and dotted magenta lines denote results of Models KK, KW, KF, and KC, respectively. 
All model results are similar in the total number of EMP survivors. 

At $\feoh \lesssim -3$, the patterns of the MDF differ. 
These extremely metal deficient stars are very early generations of stars formed with metal ejected by only one or a few SN progenitor(s) in their host halos. 
A MDF at $\feoh\lesssim-3$ is sensitive to individual SN yields. 
The MDF of Model KK has a hump at $\feoh\sim -3.6$. 
This is because iron yields of normal SNe~II are tuned to $0.07\msun$ in \citet{Kobayashi06} and metallicity of a primordial mini-halo with typical mass becomes $\feoh\sim-3.6$ by a single SN. 
We see a smaller hump at $\feoh=-4$ to $-3$ in other models, too. 
In Model KK, since energetic hypernovae blow out gas from mini-halos, the predicted number of stars with $\feoh \sim -3$ to $-2$ is smaller than other models.  

Model KC shows better consistency with observations than other models. 
The MDF of \citet{Schorck09} shows a steep drop around $\feoh = -3.6$. 
In Model KC, since iron yield and explosion energy are assumed to be the same for all the SNe~II, all the second and later generations of stars are distributed above $\feoh \sim -3.6$. 
However, theoretical iron yield strongly depend on the SN model parameters and it is difficult to distinguish which theoretical yield is the best one only from comparison with observed the MDF. 
Additionally, at such a low metallicity, the observational sample is very small and the pattern of the MDF has not yet been well revealed. 

The predicted number of stars with $\feoh \lesssim-4.5$ is very small and MDFs are bumpy because of numerical fluctuation. 
The number of observed HMP stars is comparable to the model results. 

For more metal rich stars, a MDF depends on an averaged yield but is almost independent of characteristics of individual massive stars. 
At $\feoh \gtrsim -2$, the observed MDF overwhelms the predicted number of stars by the high mass IMF models. 
One possible explanation of this excess is a changeover of the IMF. 
The IMF of metal rich stars is peaks at low mass. 
\citet{Yamada11} indicate that an IMF can be a changeover from high mass to low mass at $\feoh=-2.2$ based on statistics of stellar abundance of Zn and Co. 
If typical mass becomes lower, the number of stars surviving increases. 
At this metallicity, some thick disk stars are thought to contaminate in the sample. 
It also increases the number of stars. 
Formation of the thick disk and the IMF of these more metal rich stars will be investigated in our future works. 

\subsection{Abundance Ratio Distribution}
\begin{figure*}
\plotone{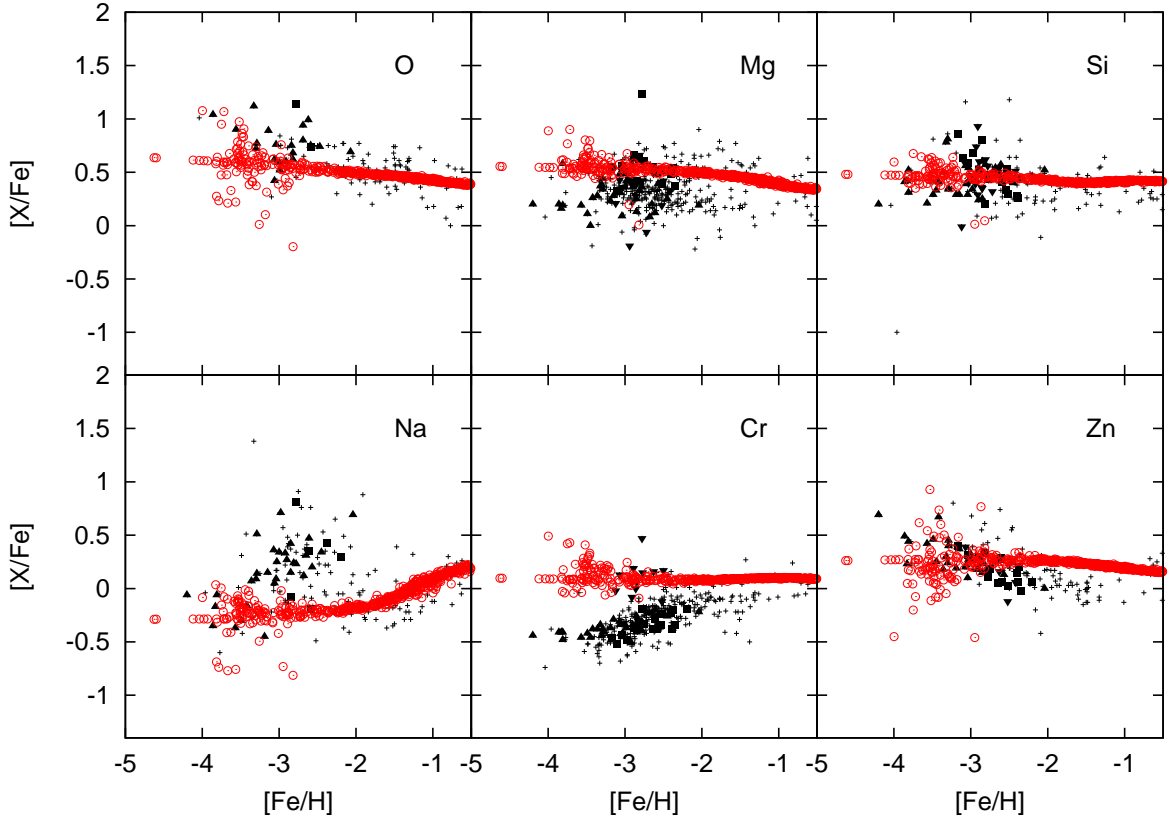}
\caption{ 
Distributions of elemental abundances of O, Mg, Si, Na, Cr, and Zn relative to iron against metallicity. 
Sampled stars in Model KK ({\it red open circles}) and observational sample compiled from the SAGA database ({\it black symbols}).  
{\it Triangles} ($\blacktriangle$): the First Stars project; 
{\it squares} ($\blacksquare$): stars analysed by Aoki et al.;
{\it inverted triangles} ($\blacktriangledown$): \citet{Honda04, Honda07}; 
{\it plus signs} ($+$): other giants compiled by the SAGA database. 
}
\label{KK}
\end{figure*}

In Figure~\ref{KK}, we show predicted and observed abundance ratio distributions for six elements for Model KK. 
The predicted typical abundance ratio and their trend against metallicity are similar to the result of the chemical evolution model by \citet{Kobayashi06}. 
But since abundance ratio differs from mini-halo to mini-halo in our model, abundance distributions show significant dispersion. 
The predicted scatters of abundances are prominent at $\feoh<-3$. 
At larger metallicity, element abundances are averaged by mixing of ejecta from many SNe and the scatters decrease. 
As shown in Paper~I, metals of stars with $\feoh=-2.5$ originate in $\sim 10$ precursory SNe in their host mini-halos. 

Most stars with $\feoh<-4$ are local first stars without any precursory SN exploded in their host mini-halos. 
Metals in these stars are originate in mixtures of matter ejected to IGM by SNe which had exploded in other mini-halos prior to the formation of the host mini-halos of these stars. 
Scatter of the abundances of these stars is smaller than the stars with $-3>\feoh \gtrsim-4$. 
Observed HMP stars show abundance anomalies of C, N, O and Na but these are thought to be due to the surface pollution by binary mass transfer \citep{Suda04, Nishimura09}. 

\subsubsection{$\alpha$-elements}
For Mg and Si, typical abundance ratio is consistent with observations. 
Observed distribution of $\abra{\alpha}{Fe}$ is almost flat against $\feoh$ at $\feoh<-1$ and the dispersion is small. 
The predicted dispersion of the $\alpha$-element abundances is comparable to or smaller than the dispersion of the observational sample. 

Observationally, $\abra{O}{Fe}$ of EMP stars shows a slight increasing trend as metallicity decreases but predicted abundances do not. 
Typical O abundance of the First Stars sample is $\sim 0.2$ dex higher than the predicted value. 
As seen in Fig.~\ref{yield}, a larger amount of O is yielded in a more massive progenitor. 
If this increasing trend is real, stars more massive than $20\msun$ thought to be the dominant source of metal in EMP stars. 
However, Mg and Si abundances show no such a trend although a larger amount of these elements is yielded in a more massive progenitor, too. 
Another possible additional source of oxygen for EMP stars is intermediate massive AGB stars. 
But AGB stars eject both O and C, O abundances of stars other than CEMP stars should not strongly affected by AGB stars. 
We note that observational determination of the oxygen abundance is difficult and there is large uncertainty on observational data. 
\citet{Nissen02} argue that, when 3D effects are taken into account, $\abra{O}{Fe}$ decreases in the metal poor stars and the increasing trend diminishes. 
For some EMP stars with low $\abra{O}{Fe}$, oxygen is undetectable. 

The sample from the First Stars project shows very small scatter for Mg and Si. 
This is consistent with the result of Model KK. 
As seen in the following, Model KK with hypernovae predicts the smallest scatters among the computations in this paper. 
It is because $\abra{O}{Fe}$ in the hypernovae ejecta is tuned to be constant and large energy of the hypernovae efficiently mix the ISM. 
The observed small scatter of the First Stars sample indicates that gas in the early universe is mixed with large mass. 
However, the subsamples by other authors show larger scatter ($\sim 0.5$ dex).
Further observations are required to understand the gas dynamics in the early universe. 
For Mg, the First Stars sample are distributed at a slightly lower $\abra{Mg}{Fe}$ than the model results but other samples are consistent. 
\citet{Andrievsky10} argue that, when non-LTE effects are taken into account, the mean value of $\abra{Mg}{Fe}$ increases and become similar to the mean value of $\abra{O}{Fe}$. 
This will improve the consistency between model and observed data. 

Observationally, it is known that $\alphafe$ decreases as $\feoh$ increases at $\feoh>-1$ for stars in the Milky Way. 
In the many chemical evolution studies, this decreasing trend has been explained by the contribution of SNe~Ia \citep[][and references therein]{Matteucci08}. 
SNe~Ia eject large amounts of iron after long delay time and decrease $\alphafe$ at large metallicity. 
Under the high mass IMF, the contribution from SNe Ia is relatively weak because the number of intermediate massive stars relative to high mass stars is small. 
In Model KK, $\alphafe$ decreases at $\feoh>-1$ but the rate of the decline is smaller than observations. 
Although the IMF is assumed to be the same for the whole metallicity range for simplicity in this study, 
our high mass IMF is derived from the statistics of the EMP survivors and the IMF should be different at higher metallicity. 

\subsubsection{Na}
For Na, observational sample of EMP stars show quite large scatter. 
All the subsamples shows large scatter and it is thought to be intrinsic. 
Observed stars with lower $\abra{Na}{Fe}$ have abundance similar to the model result but some other stars show larger $\abra{Na}{Fe}$. 
This indicates that there are additional sources of Na. 
Internal mixing may modify the surface Na abundance of some evolved EMP survivors \citep{Spite06}. 
As discussed in \citet{Nishimura09}, in intermediate massive EMP stars at AGB stage, Na can be synthesized and dredged-up by the He-flash driven deep-mixing, which is a mixing mechanism peculiar to the EMP stars. 
Stars with high $\abra{Na}{Fe}$ can be influenced by matter ejected from these AGB stars. 
At $\feoh>-2$, predicted typical abundance and their increasing trend are consistent with observations. 
We note that, when non-LTE effect taken into account, Na abundance of some stars for which high $\abra{Na}{Fe}$ value is reported  decreases to $\abra{Na}{Fe}\sim -0.2$ \citep{Andrievsky07}. 

\subsubsection{Cr}
$\abra{Cr}{Fe}$ shows a decreasing trend as decreasing metallicity. 
The predicted relative abundance is not consistent with the observations of EMP stars, as pointed out by \citet{Kobayashi06}. 
We note that the sample of \citet{Honda04} is distributed around $\abra{Cr}{Fe}=0$ and consistent with model results. 
However, the First Stars sample and Aoki's sample show clear decreasing trend with small scatter. 

\subsubsection{Zn}
In the models in this paper, Zn is mainly produced by O-Ne-Mg SNe since \citet{Wanajo09} predict a large Zn yield for O-Ne-Mg SNe. 
Hypernovae are also important sources of Zn, especially at very low metallicity. 
For EMP stars, all observational subsamples show similar abundance distribution and they are consistent with the model result. 

Observations show an increasing trend of $\abra{Zn}{Fe}$ as decreasing metallicity. 
Both subsamples by the First Stars group and Aoki et al. also show a clear increasing trend with small scatter. 
The model result shows flat distribution and predicts lower $\abra{Zn}{Fe}$ at $\feoh\gtrsim-2$. 
Recentry, \citet{Yamada11} shows that a decrease in $\abra{Zn}{Fe}$ above $\feoh>-2.2$ may be due to changeover of the IMF from high mass to low mass. 
The IMF changeover lowers the frequency of hypernova and lowers the $\abra{Zn}{Fe}$. 
The contribution from hypernovae is discussed again with results of the model without hypernovae in Section \ref{hyperS}. 

At $\feoh=-2$ to $-1$, the model predict higher $\abra{Zn}{Fe}$ than observations. 
Zn thought to be overproduced by O-Be-Mg SNe. 
Criterion to be O-Ne-Mg SNe is not yet revealed \citep{Herwig05} and the number of O-Ne-Mg SNe can be smaller. 

Predicted scatter of Zn is larger than other elements. 
This is because hypernovae and O-Ne-Mg SNe eject matter with high $\abra{Zn}{Fe}$ but normal SNe with mass $m>20\msun$ eject a small amount of Zn. 

\subsubsection{IMF dependence}

\begin{figure*}
\plotone{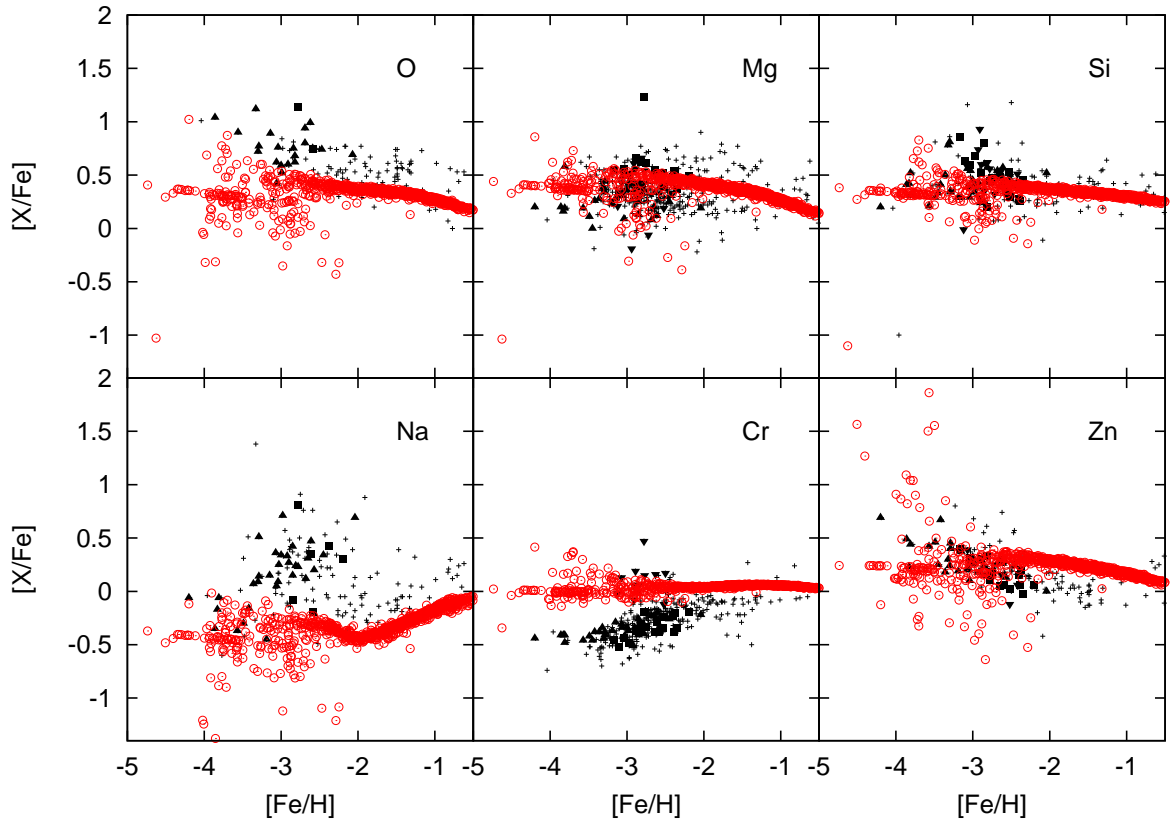}
\caption{ Same as Figure~\ref{KK} but for Model CK using low-mass IMF. 
}
\label{CK}
\end{figure*}

\begin{figure*}
\plotone{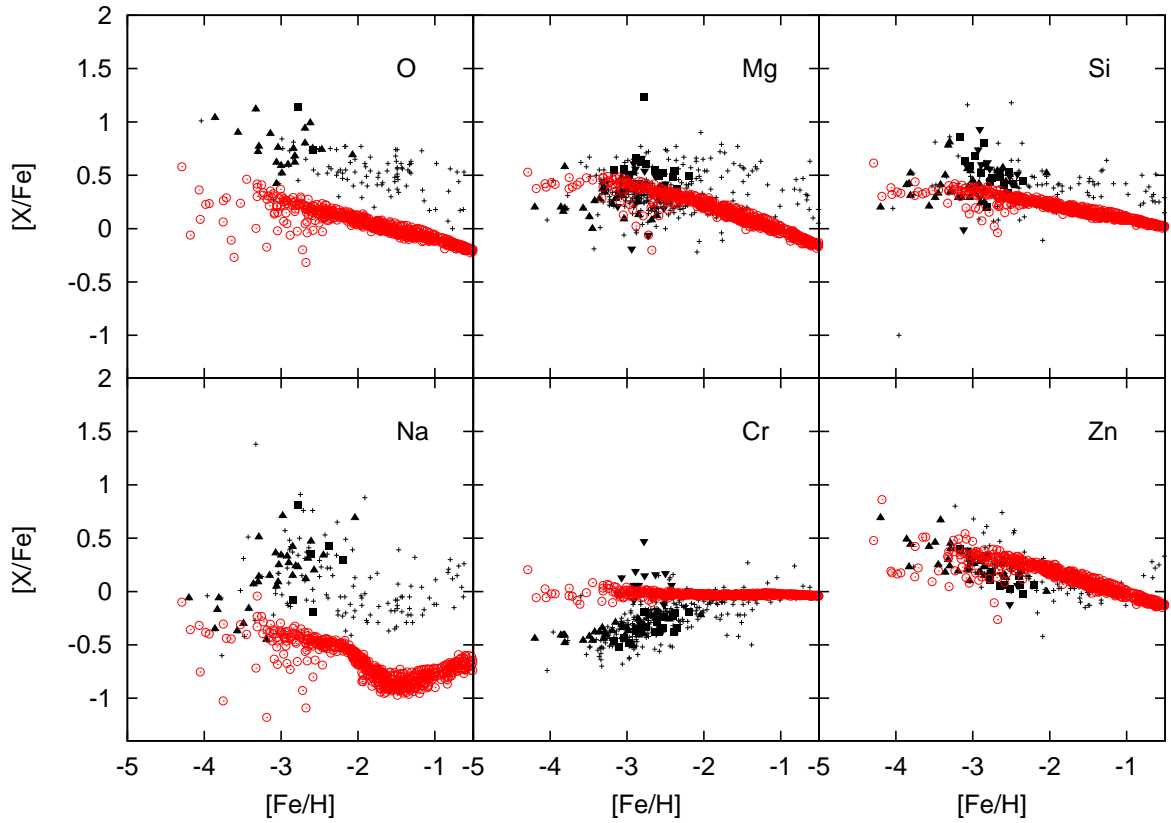}
\caption{ Same as Figure~\ref{KK} but for Model LK. 
}
\label{LK}
\end{figure*}

Figures~\ref{CK} and \ref{LK} show predicted abundance ratio distributions by models using different IMFs. 
Model CK predicts similar typical abundance ratio to Model KK. 
O, Mg and Si, are mainly provided by stars with mass heavier than $20\msun$ and iron is ejected by all SNe~II and SNe~Ia. 
For EMP stars, $\alphafe$ depends on a fraction of stars with $>20\msun$ among SNe~II and it depends on the IMF. 
For Models KK and CK, typical mass of stars is quite different but the slope of the IMFs at mass range to be SNe~II ($10-50\msun$) is similar, as seen in Fig.~\ref{IMFs}. 
Relative frequency of heavier ($>20\msun$) and lighter ($<20\msun$) SNe~II are similar for both IMFs and typical $\alphafe$ are also similar. 
However, Model CK predicts that some stars with low oxygen abundance ($\abra{O}{Fe} \sim 0.2-0.3$) are distributed at very low metallicity range ($\feoh\lesssim-3.5$). 
These stars are born in mini halos formed at low redshift. 
In this model, many SNe~Ia yield iron at lower redshift, and iron ejected from mini halos lower the $\abra{O}{Fe}$ of IGM. 
Metallicity of the IGM is still low because ejected matter is diluted in large mass. 
Mini halos formed with the IGM polluted by SNe~Ia have low $\abra{O}{Fe}$ but low metallicity. 
Observationally, these stars are not detected. 
If ejected matter is mixed in smaller mass, iron and oxygen abundance of polluted IGM become larger and these O-poor stars dissipate. 
We note that, oxygen abundance of these stars can be lower than the detection limit of O. 
Some stars without detection of O possibly have such an abundance feature. 

For Model LK, $\abra{O}{Fe}$ is obviously lower than Models KK and CK in the whole metallicity range, as seen in Fig.~\ref{LK}. 
This is because a slope of the IMF is steeper at mass range of stars to be SNe~II. 
The relative frequency of heavier SNe~II ($\gtrsim 20\msun$) is smaller and a smaller amount of $\alpha$-elements is ejected. 
The predicted distribution of $\abra{O}{Fe}$ is much lower than the observations for EMP stars. 
At $\feoh\sim -2$, the abundances of Mg, Si and Na relative to iron are also lower than the observations. 

At $\feoh\gtrsim-1$, we can see clear decreasing trends for the $\alpha$-element abundances as increasing metallicity for Model CK and LK. 
Relative numbers of SNe~Ia are larger for these models and they lower the $\alphafe$ at higher metallicity. 

\subsubsection{Supernova Models}
\begin{figure*}
\plotone{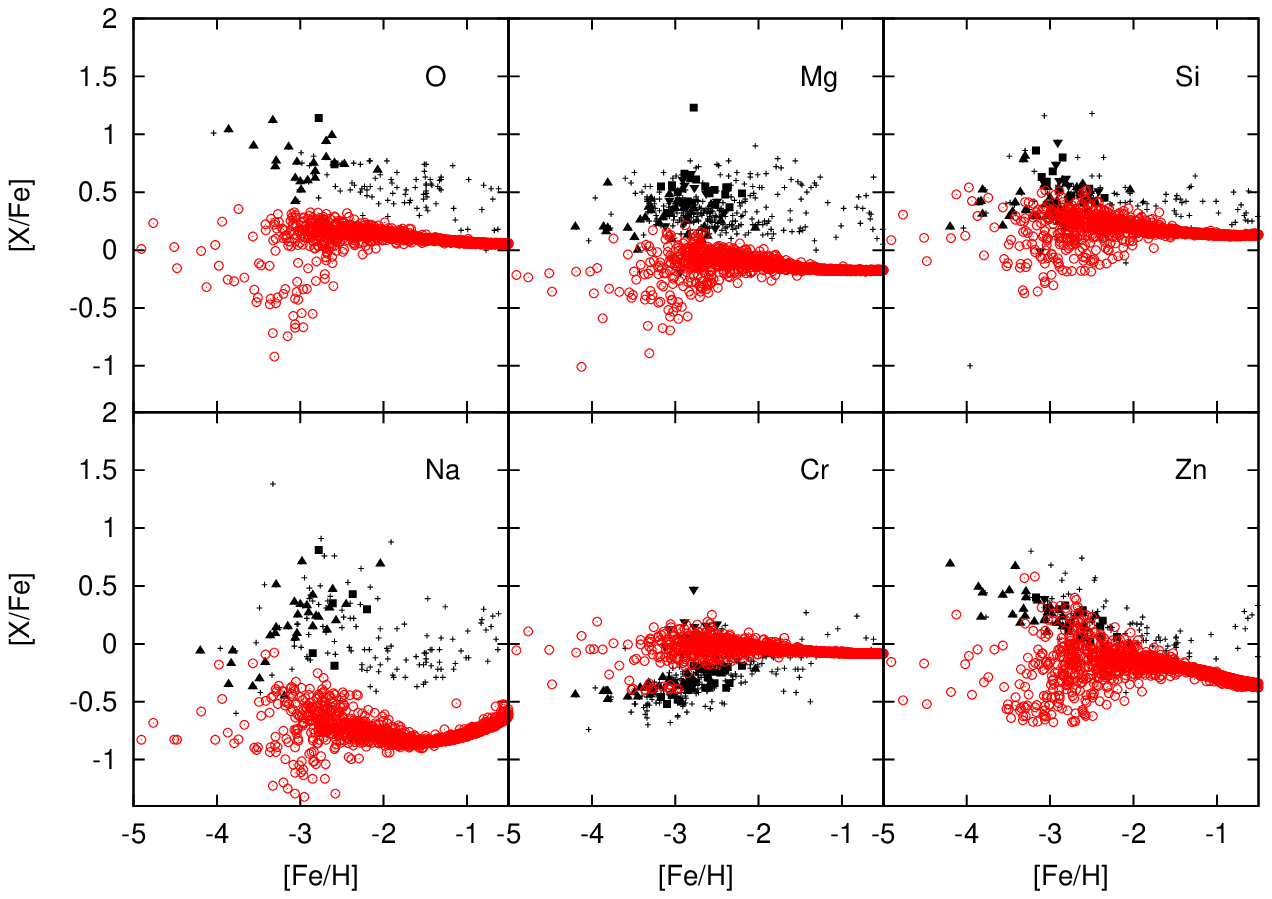}
\caption{ Same as Figure~\ref{KK} but for Model KW with SN yield by \citet{Woosley95}. 
}
\label{KW}

\end{figure*}
\begin{figure*}
\plotone{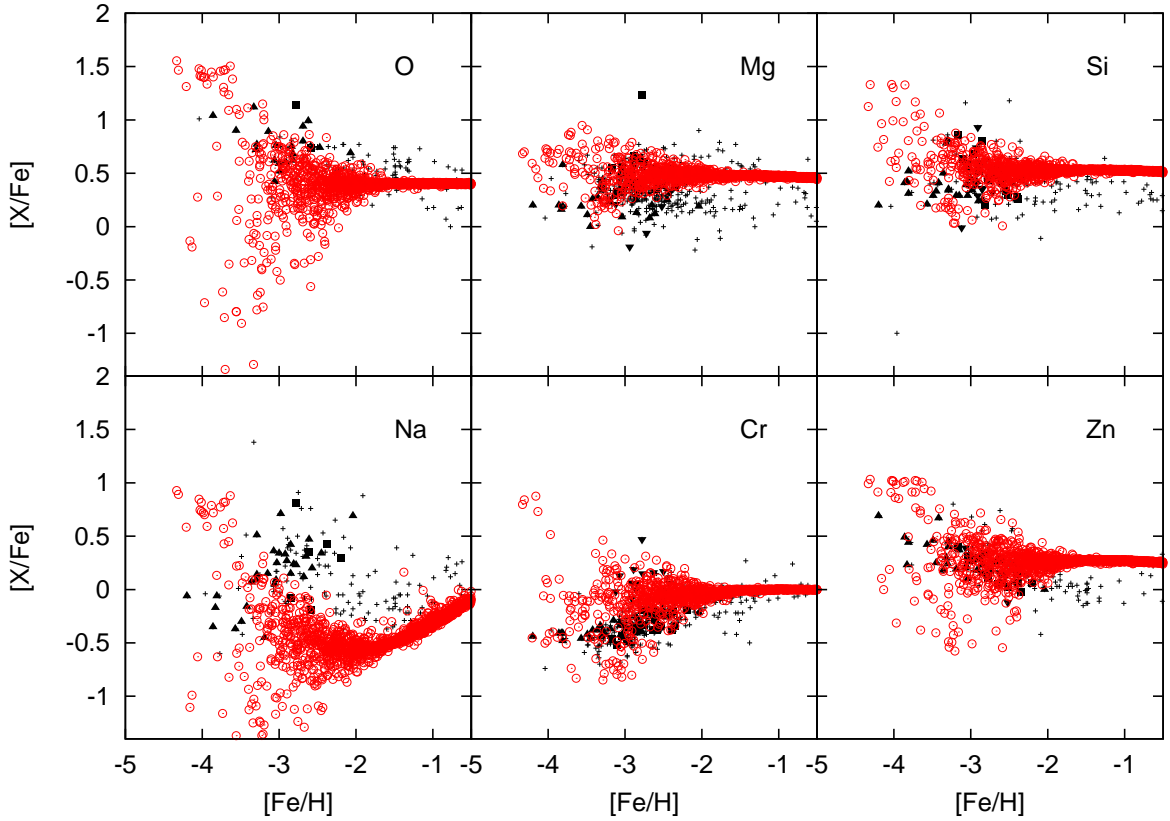}
\caption{ Same as Figure~\ref{KK} but for Model KF with SN yield by \citet{Francois04}. 
}
\label{KF}
\end{figure*}

\begin{figure*}
\plotone{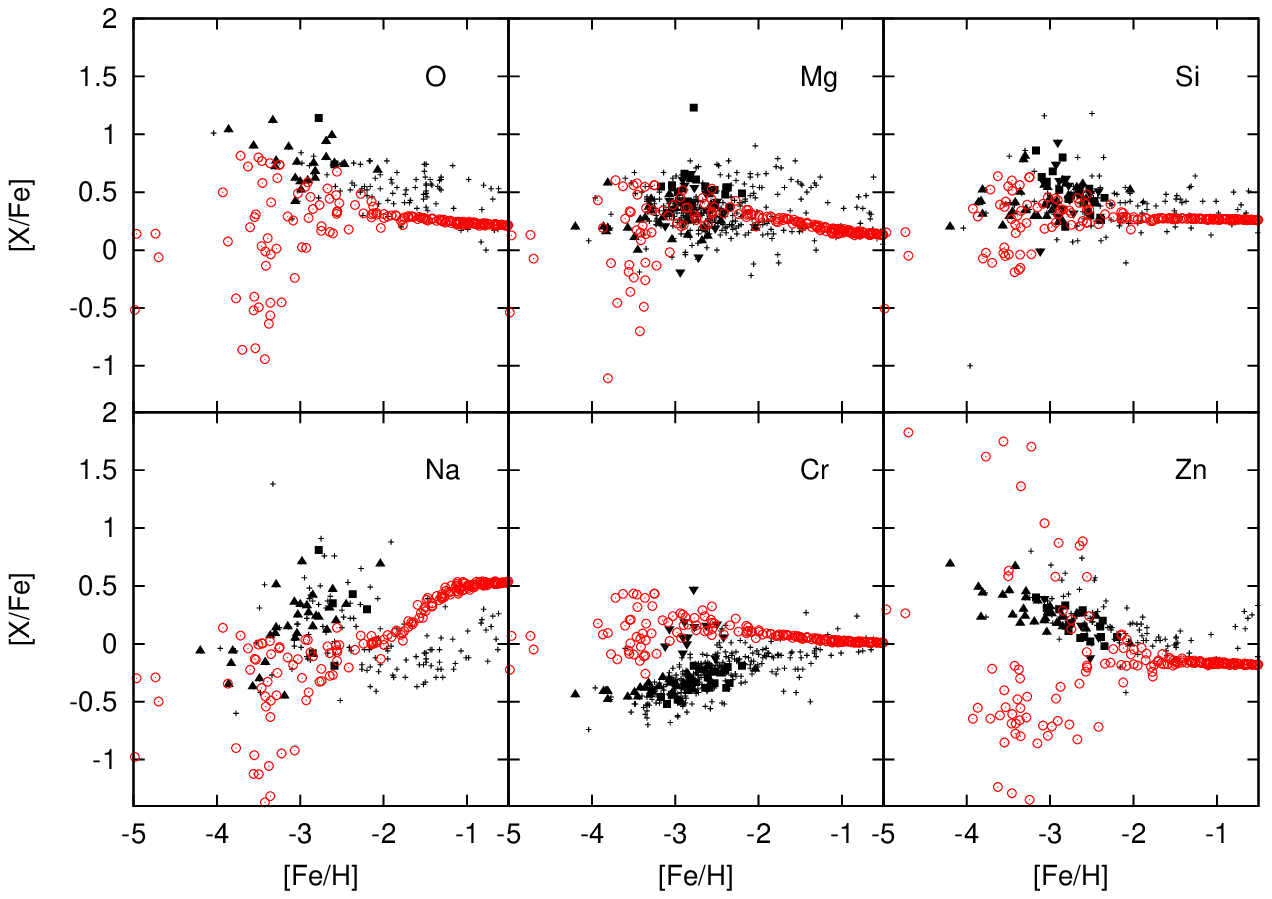}
\caption{ Same as Figure~\ref{KK} but for Model KC with SN yield by \citet{Chieffi04}. 
}
\label{KC}
\end{figure*}

Figures~\ref{KW}, \ref{KF}, and \ref{KC} show abundance ratio distributions for Models KW, KF, and KC using SN yields by \citet{Woosley95}, \citet{Francois04}, and \citet{Chieffi04}, respectively. 

For Model KW, as \citet{Francois04} pointed out, predicted $\alpha$-element abundances do not agree with observations. 
Predicted abundances of the O, Mg, and Na relative to iron are 0.5dex or more lower than the observational sample. 

For Model KF, since \citet{Francois04} modify the nucleosynthetic yields to match observational data, predicted typical abundances of EMP stars show good agreement with observations for elements other than Na. 
The decreasing trend of $\abra{Cr}{Fe}$ at low metallicity is also reproduced. 
They assume that a SN with larger initial mass ejects a smaller amount of Cr (see Fig~\ref{yield}). 
Since a star with larger mass have shorter lifetime, $\abra{Cr}{Fe}$ increases as metallicity increases with time. 
However, all other studies with nucleosynthesis computations predict that a more massive star yields a larger amount of Cr. 

While Model KF well reproduces the typical abundances, it predicts larger dispersion of the element abundances than the observations. 
Especially, $\abra{O}{Fe}$ distributes from $-1.5$ to $+1.5$ at $\feoh<-3$. 
This model also predicts some stars with $\abra{Si}{Fe}>+1$. 
The predicted scatter is much larger than the observations and it indicates that a one-zone model is inadequate to understand earliest phases of the chemical evolution and metal yields of very metal poor SNe. 

Model KC predicts typical abundances in agreement with observational sample, for Mg and Si. 
Lower O abundance than observations is predicted but observed O abundances can be lower when non-LTE and 3D effect taken into account, as mentioned above. 

This model predict some stars with very low $\abra{O}{Fe}$ and $\abra{Mg}{Fe}$ at $\feoh<-3$. 
Such abundance patterns are produced from SNe at low-mass end of the mass range to be SNe~II. 
As seen in Figure~2, stars with $10-12\msun$ yield small amount of O and Mg. 
Lower mass limit to be SNe~II is assumed to be $10\msun$ in this paper but fate of the stars with $\sim10\msun$ is not well revealed. 
Some stars with $\sim 8-12\msun$ thought to become ``super-AGB'' stars \citep{Garcia94} and evolve to O-Ne-Mg white dwarfs or electron capture supernovae \citep{Herwig05} with very little iron yield. 
Absence of the very $\alpha$-poor stars possibly indicate that the lower mass limit to be SNe~II is larger than $10\msun$ at very low metallicity. 
\citet{Kawabata09} argue that stars with $8-12\msun$ become ``faint supernovae'' with low iron yield. 
Although \citet{Chieffi04} have assumed iron yield is $0.1\msun$ for all SNe, observations indicate that some SNe yield lower amounts of iron. 

Na is overproduced at higher metallicity ($\feoh >-2$). 
Cr abundance at solar metallicity is consistent with the observations but the increasing trend is not reproduced. 

Very large scatter of Zn is predicted because they argue that a large amount of Zn is yielded in a SN~II with $m\leq13\msun$ but a very little amount of Zn is yielded in a SN with $m>13\msun$. 
When low mass limit to be SNe~II is larger as discussed above, stars with very high $\abra{Zn}{Fe}$ is not formed. 
A yield of Zn is sensitive to entropy during explosive Si-burning at SN explosion. 
As discussed later, energetic hypernovae thought to be required to explain Zn abundance of EMP stars and their trend. 

\subsection{Hypernova v.s. normal SN}\label{hyperS}
\begin{figure}
\plotone{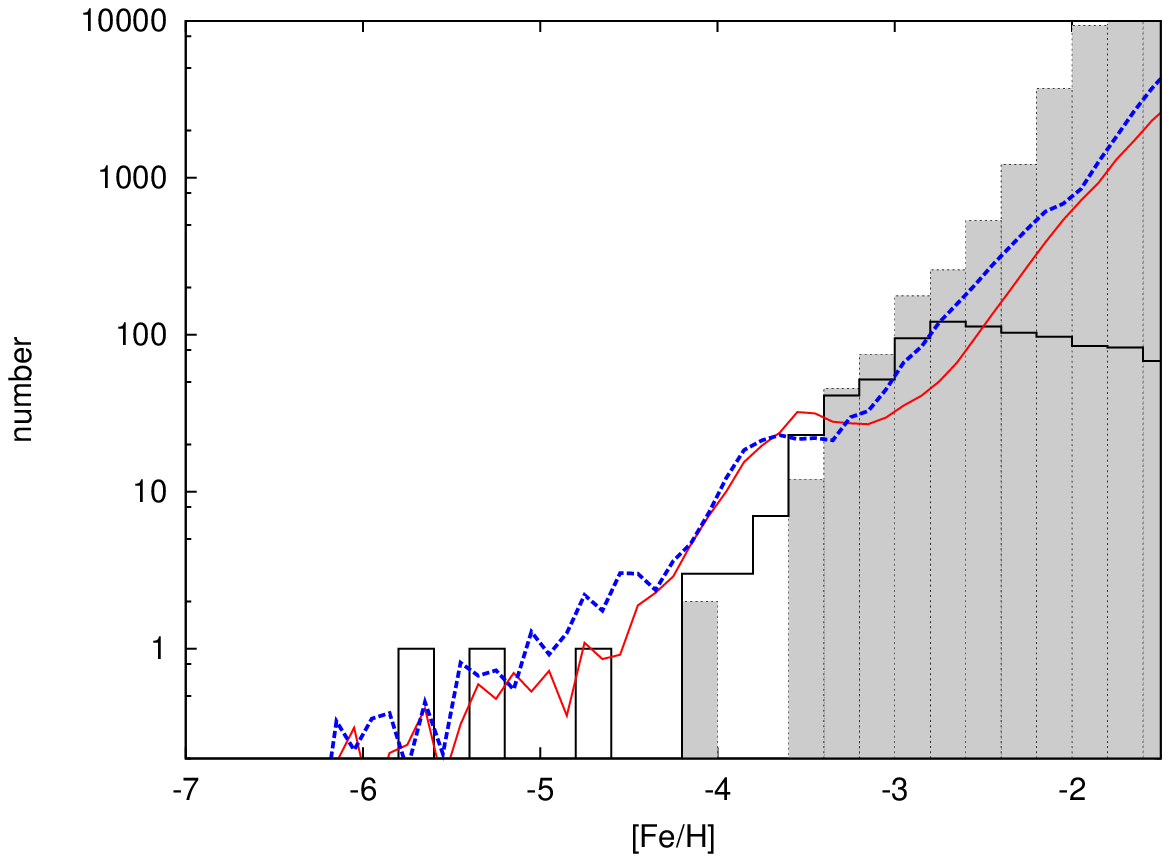}
\caption{ Effect of hypernovae on the MDF. 
The solid red and dashed blue lines denote result of Model KK and Model KKn without hypernovae, respectively.
}
\label{MDFn}
\end{figure}
\begin{figure*}
\plotone{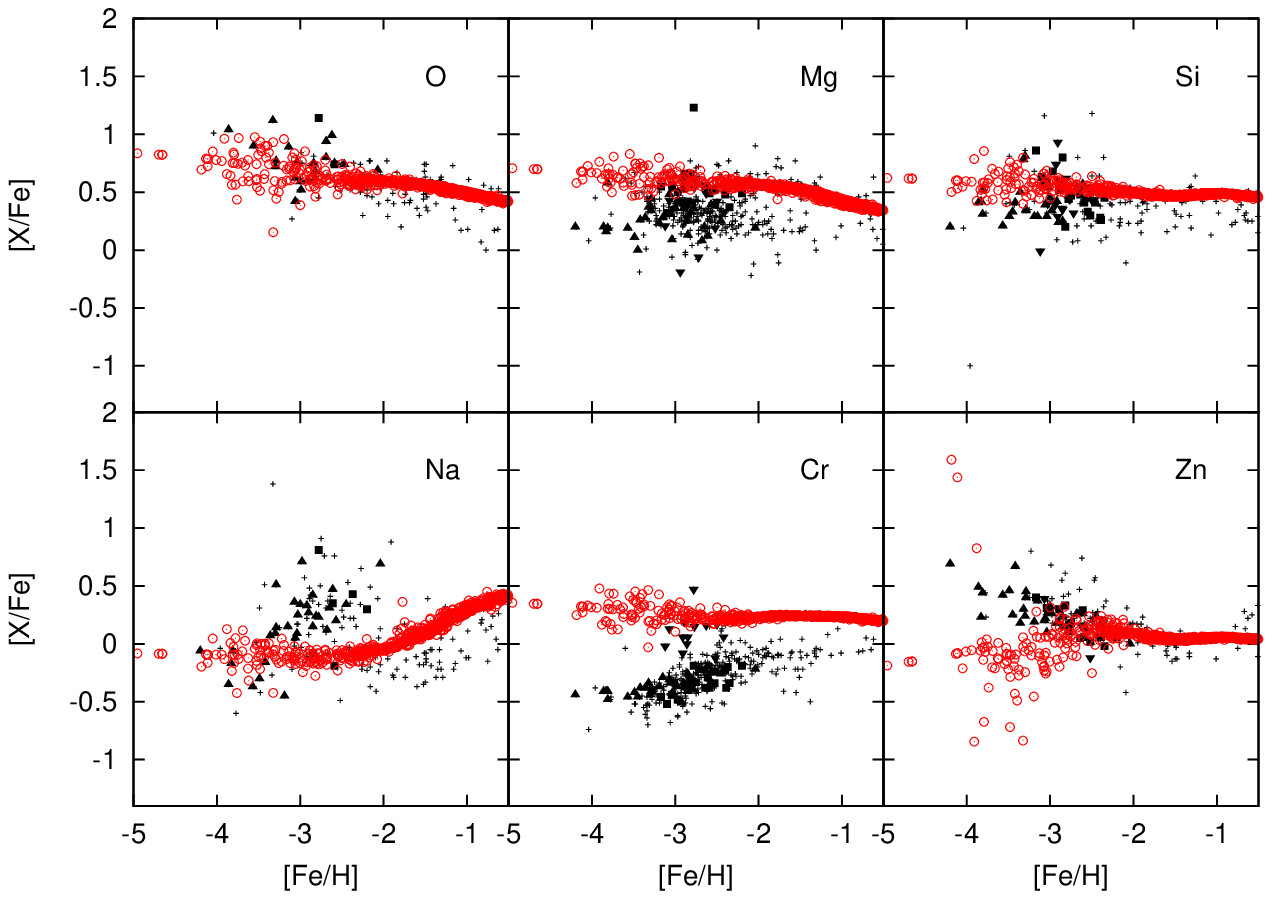}
\caption{ Same as Figure~\ref{KK} but for Model KKn without hypernovae contribution. 
}
\label{KKn}
\end{figure*}

Figures~\ref{MDFn} and \ref{KKn} show a result of Model KKn without hypernovae contribution. 
The MDF of Model KKn is similar to Model KK. 
For the abundance ratio distributions, the most plausible difference from Model KK is lower $\abra{Zn}{Fe}$. 
Since hypernovae synthesize a much larger amount of Zn than normal SNe, Zn abundance of Model KKn is lower than Model KK and lower than observations for EMP stars. 
$\alpha$-element abundances predicted by Model KKn are slightly higher than Model KK and $\abra{O}{Fe}$ shows better agreement with observations than Model KK. 
But for Mg and Cr, Model KKn predicts slightly higher relative abundances than the observations. 
This is because a normal SN yields a smaller amount of iron than a hypernova. 

Large explosion energy of hypernovae affects also gas dynamics. 
Many mini-halos are blown up by their large explosion energy and ejected metal is mixed in a large mass. 
Since it averages element abundances, the scatter of the predicted abundance of Model KK is smaller than other models. 
The observed small scatter of $\alphafe$ suggest that there were many hypernovae in the early phases of the chemical evolution. 
We note that, however, \citet{Kobayashi06} have tuned parameters in their computations to get $\abra{O}{Fe}=0.5$ for all hypernovae and the scatter of the $\alpha$-element abundances is decreased artificially. 

\section{Conclusions}\label{concS}
We compute formation history of EMP stars with a hierarchical chemical evolution model and present MDFs and abundance ratio distributions. 
We adopt various IMFs and SN yields and compare the results. 
As for IMFs, we adopt the high mass IMF given in our earlier studies\citep{Komiya07, Komiya09}, the IMF by \citet{Lucatello05}, and the standard low mass IMF by \citet{Chabrier03}. 
The pattern of the MDFs is similar for these three models but the predicted numbers of EMP survivors are quite different. 
The high mass IMF model predicts a comparable number of EMP survivors to observations but other two IMFs predict many more EMP stars in our scenario. 
Our hierarchical model reproduce a steep decline of the MDF below $\feoh=-3.6$ and a tail below $\feoh<-4$. 

Abundance ratio distributions predicted by the high mass IMF are similar to the Salpeter IMF and the predicted $\alpha$-element abundances are consistent with observations. 
Typical value of $\alphafe$ depends on the slope of the IMF at a mass range in which stars explode as SNe~II and the high mass IMF with $\mmd=10\msun$ and $\dm=0.4$ has a slope similar to the Salpeter IMF. 
The IMF by \citet{Lucatello05} which has a steeper slope at $10-50\msun$ predicts lower $\alphafe$. 

We investigate not only the averaged abundance and their trend against metallicity but also the scatter of the abundances using our hierarchical models. 
Abundance distributions strongly depend on the adopted SNe nucleosynthesis models. 
For typical abundances of $\alpha$-elements, yields by \citet{Kobayashi06}, \citet{Francois04}, and \citet{Chieffi04} show reasonable agreement with the observational sample. 
The sample from the First Stars project shows very small scatter for $\abra{Mg}{Fe}$ and $\abra{Si}{Fe}$. 
This indicates that abundance ratio ejected by various SNe is homogeneous or inter stellar matter is well mixed in large mass. 
The result with yields by \citet{Kobayashi06} is consistent with this very small scatter. 
However, stars analysed by other authors show larger scatter. 
Although the SN yields by \citet{Francois04} are the best fit yields as far as averaged abundances, their yields predict much larger scatter than all the observational sample for O and Si. 
Absence of stars with low $\abra{O}{Fe}$ and low $\abra{Mg}{Fe}$ possibly indicate that iron yields of stars with mass around $10\msun$ are lower than normal SNe II. 

Observed decreasing trend of $\abra{Cr}{Fe}$ as metallicity decreases cannot be explained by the adopted yields by the theoretical SN nucleosynthesis computations. 
For Na, intermediate massive AGB stars or internal mixing affect surface abundances of some EMP stars with Na enhancement. 
For O, Mg and Na, correction of 3D and/or non-LTE effect thought to decrease the scatter of observed abundances and improve the consistency to the model result. 
Hypernovae are the plausible dominant source of Zn for EMP stars. 
Models without hypernovae predict lower $\abra{Zn}{Fe}$ than in observations. 
Observed increasing trend of $\abra{Zn}{Fe}$ as metallicity decreases is not reproduced in our hierarchical models. 

Results in this paper are depend on the models of SN nucleosynthesis and accuracy of the abundance determination from spectroscopic observational data. 
Uncertainty of the theoretical SN yields is still large. 
Observational subsamples by different authors show different abundance distributions for some elements. 
Recent studies with 3D/non-LTE effects taken into account show that averaged abundances shift significantly by these effects and dispersion of abundances become smaller for some elements. 
Further theoretical and observational studies are required to reveal the origin of metals in the EMP stars and nature of the metal poor SNe.

\appendix
\section{Pair Instability Supernova}
\begin{figure}
\plotone{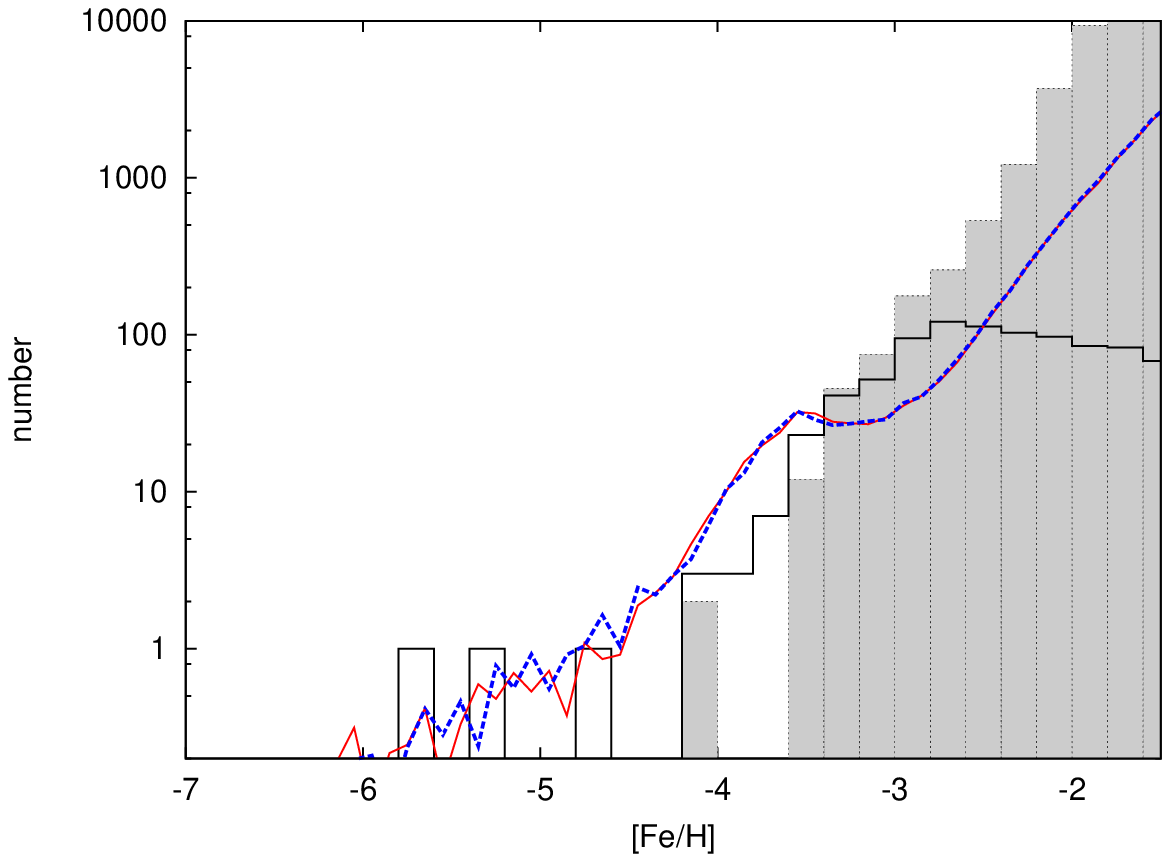}
\caption{ Same as Figure~\ref{yieldMDF} but for model with pair instability SNe. 
The solid red line denotes result of Model KK and the dashed blue line denotes result of model with pair instability SNe.
}
\label{MDFPI}
\end{figure}

\begin{figure}
\plotone{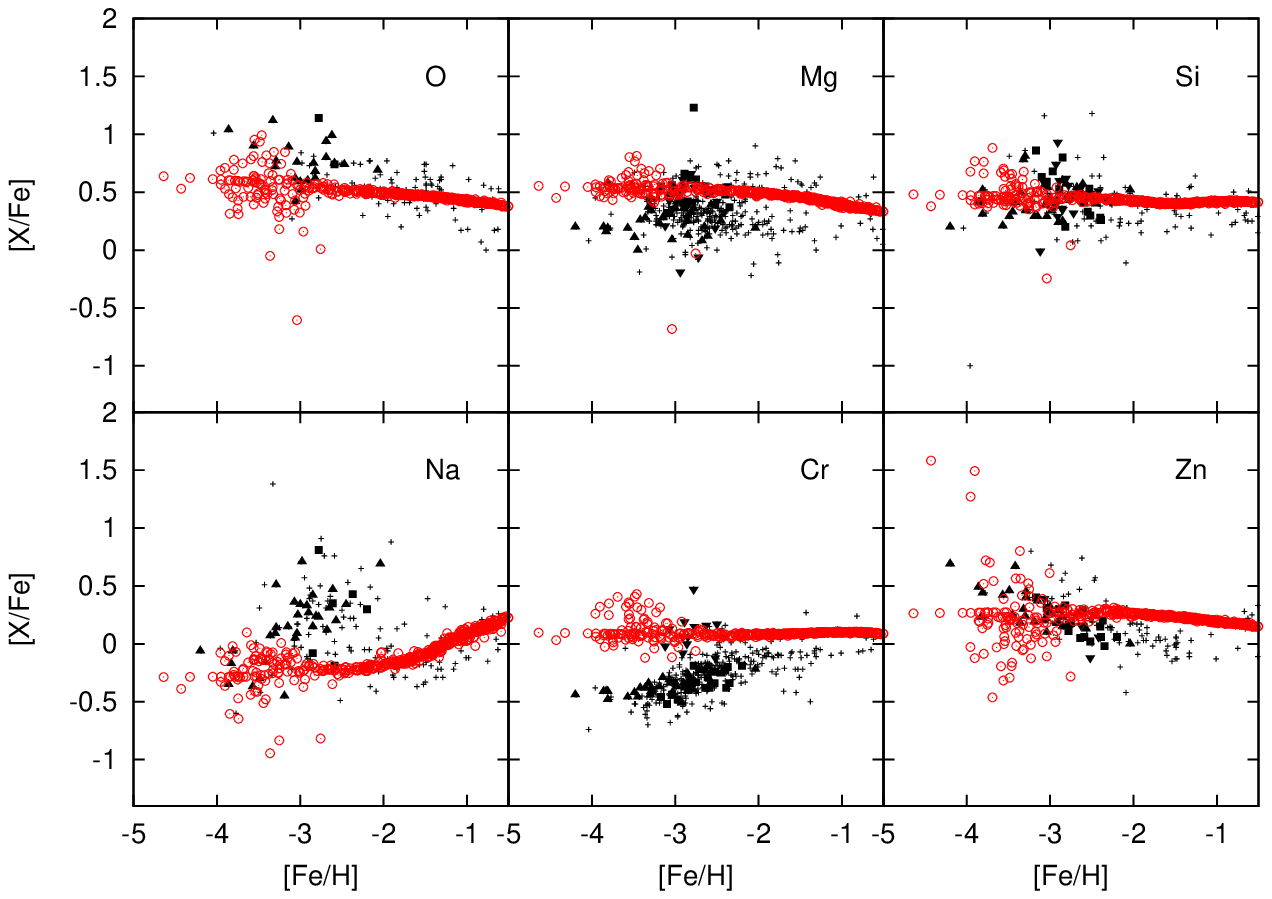}
\caption{ Same as Figure~\ref{KK} but for model with the pair instability SNe. 
}
\label{KKp}
\end{figure}

First stars formed without metal are thought to be very massive stars with mass larger than $100\msun$. 
Many numerical simulations for first star forming clouds argue that only one very massive density peak is formed without fragmentation \citep[e.g.][but see also Clark et al. 2008]{Abel02, Oshea07, Yoshida08}. 
Stars with mass $140-270\msun$ become PISNe with huge explosion energy. 
Theoretical studies of SN nucleosynthesis show that metal yields by PISNe are quite different from SNe II \citep{Umeda02, Heger02}. 
The amount of metal ejected by the energetic PISNe is much larger than from SNe II. 
They show a strong ``odd-even effect''; i.e., ratio between the yields of elements with even atomic number and odd atomic number is quite large. 
Zn production is very inefficient relative to other elements such as iron. 
However, observational studies of the element abundances of the EMP stars show that there is no nucleosynthetic signature of the PISNe. 

Metal provided by massive stars is thought to change the IMF and enable low mass star formation. 
The critical metallicity to the low mass star formation is investigated through studies of cooling processes in molecular cloud by dust and/or metal. 
Studies with dust argue that cooling by H$_2$ molecule formed on the surface of the dust sufficiently cool the gas cloud and lower their Jeans mass at metallicity $\feoh\gtrsim -6$ \citep{Omukai05, Schneider06}, and low mass stars are formed. 

In this appendix, we present a computation with PISNe taken into account. 
We assume that typical mass is very massive, $\mmd=200\msun$, for stars with $\abra{Z}{H}<-6$ and stars with mass $140-270\msun$ become PISNe.  
Other assumptions are the same as for Model KK. 
PISN yields by \citet{Umeda02} are adopted. 

Figures~\ref{MDFPI} and \ref{KKp} show MDFs and relative abundance distributions, respectively. 
The results are quite similar to the results of Model KK without PISNe and we cannot distinguish the observable signature of the PISNe. 
PISNe eject a large amount of metal at energetic explosions and the ejected metal is mixed into IGM.  
Metallicity of the IGM is enriched over $\abra{Z}{H}=-6$ by a few dozen PISNe occurring at very high redshift. 
After that, the IMF is changed to the high-mass one for EMP stars ($\mmd=10\msun$). 
Matter ejected by SNe II is mixed into the IGM and SNe II become the dominant source of the metal for stars with $\feoh\gtrsim -5$. 
Additionally, as we have shown in Paper~I, surface pollution by accretion of interstellar matter changes surface abundances of HMP stars with $\feoh<-5$ after their formation and the signature of the PISNe should be covered. 
In conclusion, the abundance pattern peculiar to the PISNe ejecta is obscured.

\end{document}